# Facile preparation of agarose–chitosan hybrid materials and nanocomposite ionogels using an ionic liquid *via* dissolution, regeneration and sol–gel transition†

Tushar J. Trivedi,[a] K. Srinivasa Rao[a] and Arvind Kumar*[b]

We report simultaneous dissolution of agarose (AG) and chitosan (CH) in varying proportions in an ionic liquid (IL), 1-butyl-3-methylimidazolium chloride [$C_4$mim][Cl]. Composite materials were constructed from AG–CH–IL solutions using the antisolvent methanol, and IL was recovered from the solutions. Composite materials could be uniformly decorated with silver oxide ($Ag_2O$) nanoparticles (Ag NPs) to form nanocomposites in a single step by *in situ* synthesis of Ag NPs in AG–CH–IL sols, wherein the biopolymer moiety acted as both reducing and stabilizing agent. Cooling of Ag NPs–AG–CH–IL sols to room temperature resulted in high conductivity and high mechanical strength nanocomposite ionogels. The structure, stability and physiochemical properties of composite materials and nanocomposites were characterized by several analytical techniques, such as Fourier transform infrared (FTIR), CD spectroscopy, differential scanning colorimetric (DSC), thermogravimetric analysis (TGA), gel permeation chromatography (GPC), and scanning electron micrography (SEM). The result shows that composite materials have good thermal and conformational stability, compatibility and strong hydrogen bonding interactions between AG–CH complexes. Decoration of Ag NPs in composites and ionogels was confirmed by UV-Vis spectroscopy, SEM, TEM, EDAX and XRD. The mechanical and conducting properties of composite ionogels have been characterized by rheology and current–voltage measurements. Since Ag NPs show good antimicrobial activity, Ag NPs –AG–CH composite materials have the potential to be used in biotechnology and biomedical applications whereas nanocomposite ionogels will be suitable as precursors for applications such as quasi-solid dye sensitized solar cells, actuators, sensors or electrochromic displays.

## Introduction

Agarose (AG) and chitosan (CH) are renewable and biocompatible biopolymers. AG is an algal polysaccharide, comprising alternating D-galactose and 3,6-anhydro-L-galactose repeating units. With its excellent ability to form thermo-reversible gels in hot water, AG finds numerous applications, which include food industry, pharmaceutical formulations, electrophoresis, tissue engineering or as a matrix for soft-matter organic devices.[1,2] On the other hand, CH is a linearly linked polysaccharide derived from natural biopolymer chitin, and composed of randomly distributed (1–4)-linked D-glucosamine (deacetylated unit) and *N*-acetyl-D-glucosamine (acetylated unit). Because of its biodegradable, biocompatible, non-toxic, antimicrobial and metal-binding properties, CH has been widely studied in chemical, biochemical and biomedical fields, and is extensively used in pharmaceutical and biomedical fields for controlled release of drugs, wound management, space filling implants, *etc.*[3–5] Since CH is a good interacting polymer having amine groups, it can combine or interact with an AG hydroxyl group to form an AG–CH complex, forming environmentally safe composite materials with combined advantages suitable for different applications.[6–8] However, the preparation of composites of these biopolymers in water or common organic solvents under mild conditions is not an easy process due to the presence of many hydroxyl groups, intra- and intermolecular hydrogen bonding and molecular close chain packing which creates difficulties in their effective solubilization. Besides solubility limitations, there are several other problems, such as solvent handling, volatility, generation of poisonous gases or waste and solvent recovery generally encountered while processing the biopolymers in common organic solvents. Therefore, considering the green chemistry concern, sustainability and eco-efficiency, it is imperative to use greener and high solvating ability solvents for developing novel and efficient processes of composite formation.

[a]*AcSIR, CSIR-Central Salt and Marine Chemicals Research Institute, G. B. Marg, Bhavnagar-364002, India*
[b]*Salt and Marine Chemicals Division, CSIR-Central Salt and Marine Chemicals Research Institute, G. B. Marg, Bhavnagar-364002, India.*
*E-mail: mailme_arvind@yahoo.com, arvind@csmcri.org*
†Electronic supplementary information (ESI) available. See DOI: 10.1039/c3gc41317a

Ionic liquids (ILs) (salts with melting points <100 °C[9]) normally comprise large asymmetric organic cations and inorganic or organic anions of varying sizes, and due to their unique accessible physicochemical properties, ILs are fast replacing non-volatile hazardous conventional organic solvents in different chemical and biochemical processes.[9–14] Because of the high solvating ability for biomaterials, ILs have been extensively explored for the deconstruction of biomass, particularly lignocellulosic and cellulosic materials.[15–21] ILs as cellulose dissolving solvents have been critically reviewed in Tom Welton's research group very recently.[22] Apart from cellulosic materials, the dissolution/regeneration of chitin/CH or AG has also been carried out in ILs, and useful materials have been prepared from these polysaccharides[23–31] (Note: there are many other biomaterials which have been treated with ILs but not referred to here as the present study intends to focus on AG and CH). In view of the ease of their processing in ILs, the preparation of advanced functional biopolymer composites of chitin/CH with cellulose, carbon nanotubes, or silk for different applications, such as biodegradable biosorbents, films, membranes, fibers, actuators, and composite hydrogels for cell adhesion and growth has been reported using IL technology.[32–36] In contrast, the composites of AG in ILs have been rarely studied,[37] and no reports are available on the preparation of composites of AG with CH using ILs, which can have several advantages over the conventional processes. Besides composite preparation, it is also possible to confine ILs as gel matrices by cooling biopolymer–IL solutions at ambient temperatures to form "ionogels" which are shown to be smart polymeric conducting materials that combine the chemical versatility of an ionic liquid with the morphological versatility of a biopolymer.[30,31,38–47] Such ionogels have been found to be suitable for an array of electrochemical applications.[38,46] Therefore, in addition to the preparation of AG–CH composites, we have attempted to prepare AG–CH–IL composite ionogels using a high solvating ability IL, [$C_4$mim][Cl], as the dissolution medium.

It is reported that the versatility of composites or composite ionogels for different applications can be tremendously improved by immobilizing metal nanoparticles.[48–50] In particular, the inclusion of silver oxide nanoparticles (Ag NPs), because of very good optical/conducting,[51] catalytic,[52] sensing[53] or antimicrobial properties,[54,55] can be very useful for improving the mechanical strength, conductivity and antimicrobial properties of composites and ionogels. Hence, we decided to produce Ag NPs –AG–CH composites and ionogels. Normally the nanocomposites are prepared either by incorporating nanoparticles into the composite solutions externally or by synthesizing the nanoparticles *in situ*. The latter technique is always preferable as it avoids the use of any external reducing agent and capping agent.[56,57] As both the CH and AG have the ability to act as reductants as well as stabilizers,[58–60] and can be solubilized in [$C_4$mim][Cl], we could synthesize Ag NPs in AG–CH–IL solution *in situ* and produce Ag NPs decorated nanocomposites and ionogels with ease in a single step.

In brief, the manuscript presents a simple and viable methodology for AG–CH composite and Ag NPs –AG–CH nanocomposite formation through dissolution and regeneration using [$C_4$mim][Cl]. A detailed structural, conformational and morphological characterization of composites and nanocomposites is provided. Composite and nanocomposite ionogels have been prepared and characterized for gelling, melting, rheological and conducting properties.

## Experimental

### Materials

IL, 1-butyl-3-methylimidazolium chloride [$C_4$mim][Cl], with stated purities higher than 98% mass fraction, agarose (Type II-A) and chitosan (medium molecular weight) were purchased from Sigma Aldrich. IL was dried under vacuum prior to use. Agarose and chitosan were used as received. Silver nitrate ($AgNO_3$) with a purity of 98% mass fraction was purchased from SD FINE CHEM, Mumbai. The molecular structure of IL and repeating units of agarose and chitosan is shown in Fig. 1.

### Methods

**AG–CH composite and composite ionogels.** Preparation of AG–CH composite in IL was carried out in 10 ml beakers with continuous magnetic stirring. A total amount of 5 wt% of agarose and chitosan in varying proportions (100 : 0, 80 : 20, 60 : 40, 50 : 50, 40 : 60, 20 : 80, 0 : 100) was added in pre-heated [$C_4$mim][Cl] at 100 ± 1 °C. For example, for the preparation of AG–CH (50 : 50) composite, 5 g IL was preheated at 100 ± 1 °C and then 0.125 g of powdered AG was added with continuous stirring until dissolution. After the AG was completely dissolved, we added 0.125 g of powdered CH to the resulting AG–IL solution and carried out the dissolution process till a homogeneous solution was obtained. To check the repeatability of the dissolution process, experiments were carried out at least three times. The dissolution process was monitored visually. AG–CH composite materials were either regenerated from the AG–CH–IL solutions by adding anti-solvent methanol or composite ionogels were obtained by cooling AG–CH–IL solutions at room temperature. Complete regeneration was ensured by the addition of an excess amount of methanol to the solutions remaining after regeneration. No further precipitation indicated complete regeneration. Regenerated composite material was dried in an oven at 70 °C and used for the physicochemical characterization. A total amount of 0.238 g of dried composite material was obtained and the yield was determined by taking the ratio of regenerated and original materials. The yield of the regenerated material was always >95 wt%. After regeneration of composite materials, IL was recovered from the methanolic solutions using a rotary evaporator and reused for further experimentation.

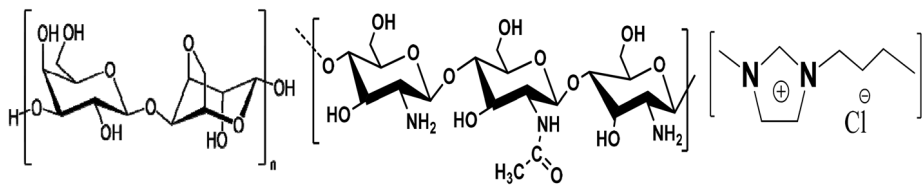

Fig. 1 Repeating structure of (left) agarose, (middle) chitosan, and (right) molecular structure of the IL, 1-butyl-3-methylimidazolium chloride.

**Ag NPs–AG–CH nanocomposite and nanocomposite ionogels.** For the preparation of nanocomposites, 100 mM silver nitrate was added into an AG–CH (50:50)–IL solution. The mixture was heated at 100 °C under continuous stirring for 6 h. The appearance of colour changes from slight yellow to dark yellow to light brownish confirmed the formation of silver oxide nanoparticles *in situ*. Addition of methanol to the resulting solution enabled precipitation of nanocomposite materials, whereas cooling the resulting solutions produced nanocomposite ionogels.

### Characterization

The molecular weights of native or regenerated AG, CH, and their composites were determined by high temperature gel permeation chromatography (HT-GPC) using a Waters 2695 Separation Module equipped with a 2414 RI detector and having Ultrahydrajel 500 and 120 columns in series. Columns were eluted with 0.1 M aqueous $NaNO_3$ at a flow rate of 0.5 ml $min^{-1}$. Calibration was performed using a dextran standard ranging from 401 000 to 4400 peak molecular weight. The concentration of AG and CH/AG–CH composite solutions was 0.02 wt% in water and 1% acetic acid solution respectively. FT-IR spectra of the native and regenerated biopolymers and composite materials were recorded at room temperature using a NICOLET 6700 FT-IR spectrometer. Conformational analysis was performed by recording circular dichroism (CD) spectra of AG and CH/AG–CH composite solutions (0.05 wt%) in water and 1% acetic acid respectively in the wavelength range of 180 to 240 nm using a Jasco J-815 CD spectrometer. Experiments were carried out in a 1 cm path length cuvette at 25 °C, and were expressed as the average of five scans. The response time and the bandwidth were 2 s and 0.2 nm respectively. Samples for recording the spectra were taken in a quartz cuvette which was immediately sealed after sampling to avoid evaporation. The desired temperature was achieved with an in-built peltier device.

The melting or denature temperature ($T_D$) and decomposition temperature ($T_{dec}$) for native, regenerated and composite materials were determined by differential scanning calorimetric (DSC) measurements using a Mettler Toledo DSC822 thermal analyzer. Measurements were performed between 30 and 450 °C at a heating rate of 5 °C $min^{-1}$. Thermogravimetric analyses (TGA) were performed using a TGA/SDTA851 Mettler Toledo under a nitrogen atmosphere from 30 to 450 °C with a heating rate of 10 °C $min^{-1}$. The surface morphology of composites was examined by scanning electron microscopy (SEM) using a LEO 1430 VP Carl Zeiss scanning electron microscope. For recording the micrograph, samples of equal thickness were gold-coated.

The melting and gelling temperatures of ionogels were determined following the method described by Craigie & Leigh.[61] The gel strength (g $cm^{-2}$) was measured using a Nikkansui type gel tester (Kiya Seisakusho Ltd, Tokyo, Japan). Measurements were performed on 3 wt% AG, CH, AG–CH and Ag NPs–AG–CH ionogels using a solid cylindrical plunger 1.127 cm in diameter. Viscoelastic measurements of various ionogels and nanocomposite ionogels were carried out on a rheometer (RS1, HAAKE Instruments, Karlsruhe, Germany). The cone/plate (60 mm diameter, 1° rad angle) geometry was selected for dynamic viscoelastic measurements. The frequency dependences of dynamic storage ($G'$) and loss ($G''$) moduli were examined in the linear viscoelastic regime (predetermined at each temperature). The oscillation measurements (temperature dependence) of $G'$ and $G''$ of ionogels were carried out in the controlled deformation mode with a strain value of 1% and a frequency of 1 rad $s^{-1}$, the temperature of the gel being maintained at 25 °C using a DC 50 water circulator. The steady-state current–voltage curve of the ionogel electrolyte was determined in the voltage range between −5 V and 5 V at a scan rate of 0.5 mV $s^{-1}$ using a source meter unit, KETHLEY model 2635A.

The formation of silver oxide nanoparticles (Ag NPs) was monitored by UV-Vis spectroscopic measurements using a spectrophotometer (Cary 500, Varian). Ag NPs in nanocomposites were visualized by SEM and transmission electron microscopy (TEM). For TEM, the samples were prepared by putting a drop of dispersed nanocomposite solution on the carbon-coated copper grid (300 mesh). The residual liquid was blotted immediately. Samples were imaged under a JEOL JEM-2100 electron microscope at a working voltage of 80 kV. The XRD measurements were performed using a Philips X'pert MPD system with CuKα radiation ($\lambda$ = 1.54056 Å) at a scan rate of 2° $min^{-1}$ with a step size of 0.03.

## Results and discussion

### AG–CH composites

Composites of varying AG–CH ratios were constructed from their homogeneous solutions prepared in [$C_4$mim][Cl] by adding the antisolvent methanol. The molecular weight and polydispersity index of the regenerated individual polymers and composite materials are listed in Table 1. Results show that with an increase of the CH proportion in the composite,

Table 1 Molecular weight ($M_n$ and $M_w$), polydispersity index (PDI), dehydration ($T_D$), and decomposition temperature ($T_{dec}$) from DSC and $T_{dec}$ from TGA measurements of native, regenerated and composite materials

| Pure/composite materials | $M_n$ | $M_w$ | PDI | DSC | | TGA |
| --- | --- | --- | --- | --- | --- | --- |
| | | | | $T_D$ | $T_{dec}$ | $T_{dec}$ |
| AG | 83 950 | 163 000 | 1.83 | 98.79 | 269.5 | 268.2 |
| Reg-AG | 78 760 | 162 348 | 1.93 | 111.75 | 270.3 | 269.0 |
| AG–CH (80/20) | 77 938 | 171 789 | 2.18 | 100.10 | 271.0 | 270.9 |
| AG–CH (60/00) | 76 214 | 184 752 | 2.41 | 96.84 | 272.1 | 271.9 |
| AG–CH (50/50) | 77 896 | 221 890 | 2.72 | 107.49 | 274.2 | 273.5 |
| AG–CH (40/60) | 78 452 | 245 877 | 2.74 | 107.80 | 283.1 | 270.4 |
| AG–CH (20/80) | 88 315 | 258 192 | 2.82 | 89.29 | 284.3 | 287.3 |
| Reg-CH | 97 442 | 283 191 | 2.91 | 92.41 | 291.4 | 292.8 |
| OCH | 98 382 | 286 020 | 2.93 | 85.37 | 305.8 | 308.8 |

AG: agarose; CH: chitosan.

the average molecular weight and PDI increased proportionally indicating no degradation of either the individual polymers or composites while being regenerated from the IL solution. The possible composite or complex formation in the material was examined from the FTIR spectra of various samples (Fig. 2, left). FTIR spectra of individual biopolymers and regenerated AG–CH composites were evaluated on the basis of specific bands and molecular motions exhibited by different functional groups. The spectrum of native CH shows the following: a broad band at 3200–3500 cm$^{-1}$ is attributed to the –NH$_2$ and –OH stretching vibrations, a peak at 1560 cm$^{-1}$ is for the NH bending (amide II) (NH$_2$), 1647 cm$^{-1}$ is due to the C=O stretching (amide I) O=C-NHR, 2927, 2884, 1411, 1321 and 1260 cm$^{-1}$ correspond to CH$_2$ bending due to the pyranose ring, 1078 cm$^{-1}$ is for saccharide structures and the band at 1380 cm$^{-1}$ is due to CH$_3$ wagging.[62] FTIR spectrum of AG shows the following: an absorption band at about 3400 cm$^{-1}$ is associated with O–H stretching, bands at 773, 894, and 932 cm$^{-1}$ are because of 3,6-anhydro-L-galactose skeletal bending, and a band at 1072 cm$^{-1}$ is attributed to the deformation mode of the C–O groups.[63,64] As can be seen in Fig. 2a, the spectra of composites show a combination of functional groups originating from both CH and AG. An increase in the AG content in the composite decreases the intensity of the band arising from the NH bending (amide II) at 1560 cm$^{-1}$ and increases the band absorbance at 1380 cm$^{-1}$. The peak at 1647 cm$^{-1}$, attributed to the C=O stretching (amide I) O=C-NHR of CH, is shifted towards lower frequency as the concentration of AG in the composite is increased. A gradual shift of the characteristic absorption bands of CH and AG indicates AG–CH complex formation either by the formation of hydrogen bonds between the –OH/–NH$_2$ groups in CH molecules and the –OH groups in AG or by the reciprocal entanglement between the macromolecular chains.[65]

The expected conformational changes while processing the biopolymers in IL were examined by comparing the circular dichroism (CD) spectra of native AG, CH, or regenerated AG, CH and AG–CH composite materials at different proportions. In the case of AG, a characteristic CD spectrum with a positive band centered at ~190 nm is observed, whereas in CH a negative band centered at around ~210 nm is observed.[30,66] Analysis of the CD spectra in Fig. 2 (right) reveals that AG and CH regenerated from IL solutions largely maintain their native confirmation, whereas in AG–CH composites, the spectra are red shifted with a decrease in intensity from positive to negative, i.e. towards the native CH with an increase in the CH amount in the composite indicating the disruption of the ordered structure of AG due to complexation/intermolecular interactions. CD spectra of AG–CH (50 : 50) gave equal intensity in positive and negative bands showing a homogeneous blending of the biopolymers in IL.

Thermal properties of composites were determined by DSC and TGA measurements. DSC thermograms of both native and AG–CH composite materials exhibited a broad endothermic peak at 85 to 110 °C (Fig. 3). Similar endothermic peaks are reported for AG–CH composites in the literature.[8] These peaks correspond to the dehydration temperature ($T_D$) of the polymers and indicate the strength of water–polymer interaction.[24,67] The variation in the position and peak area for various composites suggests the physical and molecular changes due to the interactions between AG and CH. Sharp exothermic peaks around 270 to 305 °C show the decomposition temperature ($T_{dec}$) of native and composite materials (Fig. S1†). Exothermic peaks are shifted towards higher values (red shift) when the CH content is increased in the composite indicating enhanced thermal stability of AG–CH as compared to native AG, and confirm that the interaction between the hydroxyl groups of agarose and the amino groups of CH established in the IL solution remained in the regenerated material. $T_D$ and $T_{dec}$ temperatures are recorded in Table 1. TGA profiles (Fig. 3, right) show that the decomposition of native and AG–CH composite materials took place in two stages in which the first transition occurred below 100 °C due to weight loss of samples by moisture vaporization and the second transition in a narrow temperature range from 265 to 310 °C (Table 1) which closely matches the DSC profiles for water evaporation and thermal degradation temperature of native and composite materials. The morphology of regenerated and composite materials examined by SEM micrographs (Fig. 4) shows that the surface of regenerated individual polymers resembles native polymers and the composites show features of both AG and CH showing a homogeneous blending. CH shows a smooth and homogeneous surface with some straps while the AG surface shows roughness. The surface roughness of composites decreases when the CH content increases in the composite and becomes very smooth at high CH concentrations.

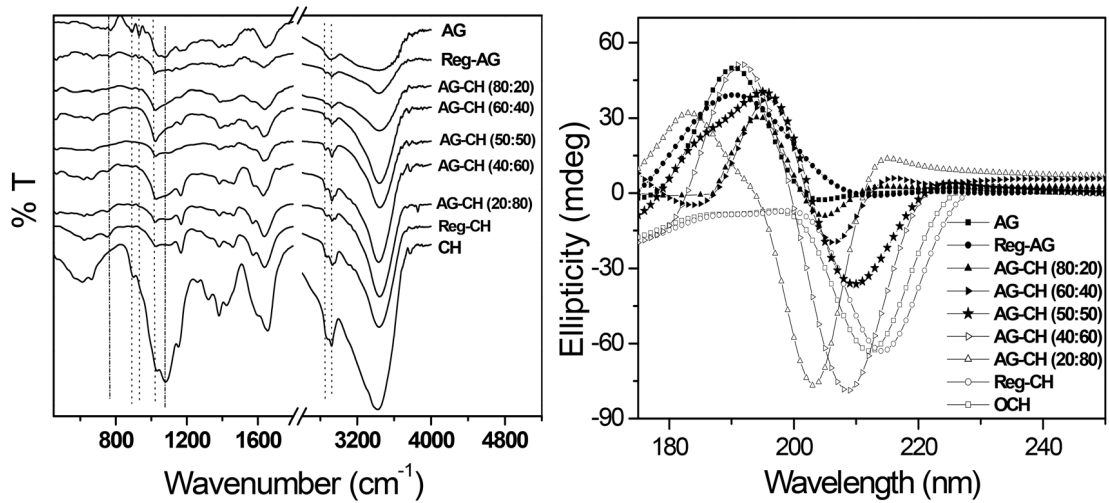

Fig. 2 FTIR (left) and CD spectra (right) of native, regenerated biopolymers, and agarose–chitosan (AG–CH) composites having different proportions.

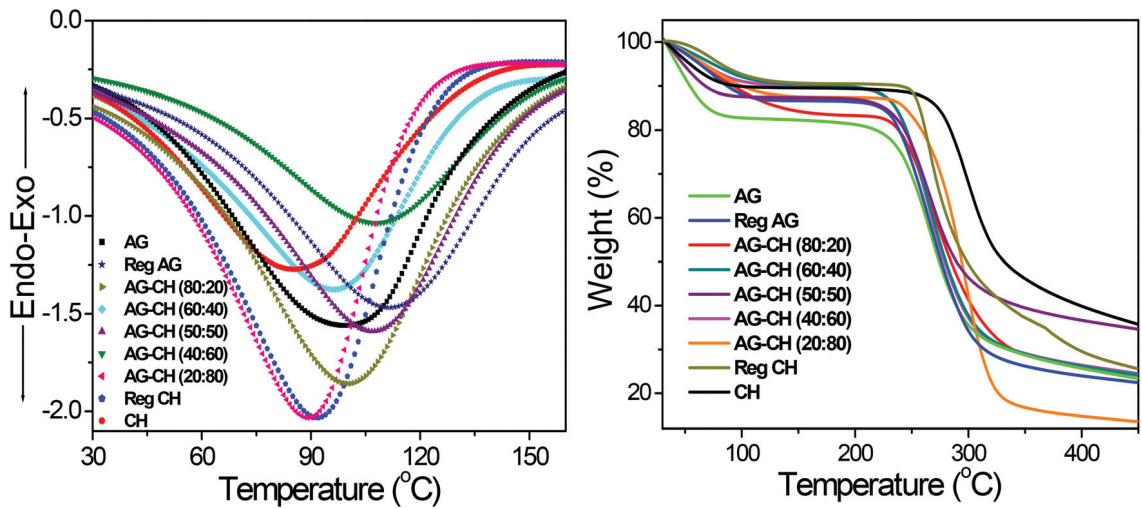

**Fig. 3** DSC ($T_D$) (left) and TGA (right) profiles of native, regenerated biopolymers, and agarose–chitosan (AG–CH) composites having different proportions.

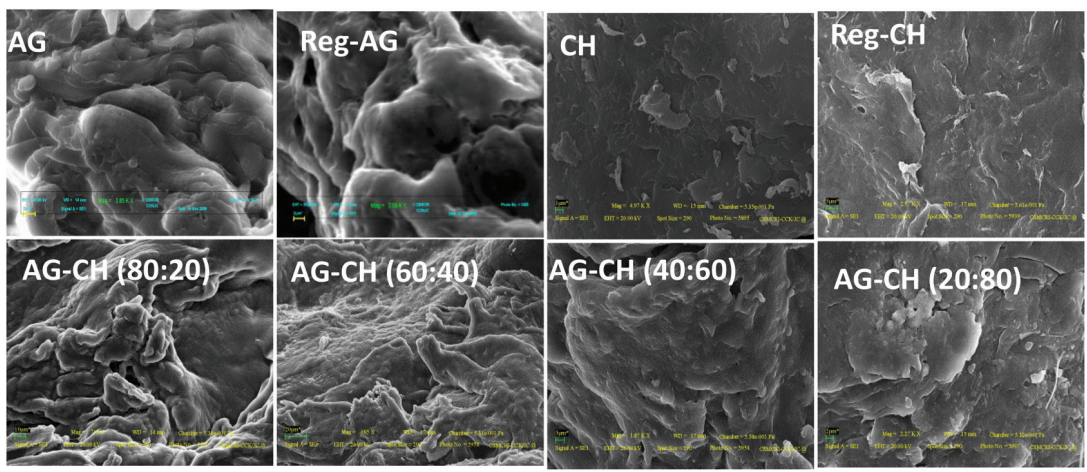

**Fig. 4** Typical SEM micrographs of native, regenerated biopolymers, and agarose–chitosan (AG–CH) composites having different proportions.

### AG–CH composite ionogels

Upon cooling, the AG, CH or AG–CH (50 : 50) dissolved viscous solutions of IL resulted in the formation of thermoreversible conducting ionogels (Fig. 5). For a reasonable comparison we prepared 3 wt% ionogels. Dried and powdered AG, CH or AG + CH was dissolved in preheated [C$_4$mim][Cl] and allowed to cool till gelled. Gelling and melting temperatures of various ionogels were monitored through visual inspection of gel and liquid state and are noted in Table 2. Rheological measurements (discussed later) also provided melting temperatures which were very close to that observed by visual inspection. The gel strength measured using a Nikkansui type gel tester at 30 °C is also given in Table 2. Like AG hydrogels, all the ionogels except that of pure AG-[C$_4$mim][Cl] showed thermal hysteresis. The formation of ionogels is expected to occur in a slightly different fashion from the formation of water based gels. In the case of hydrogels the AG gelation in water is promoted by coordinated molecular conformational changes at low temperatures; however, as indicated by the very small hysteresis, this mechanism is not followed in [C$_4$mim][Cl]. In the case of AG, which is essentially a neutral biopolymer, hydrogen bonding between IL ions and hydroxyl groups of AG molecules must be playing an important role in gelling. The presence of a large organic cation will interact with AG strands and prevent the formation of double helical structures, and finally lead to a comparatively weaker gel. On the other hand, CH does not produce temperature induced hydrogels, because of the relatively rigid molecular conformation given by the beta glycosidic links of chitosan which are further stabilized by hydrogen bonds among consecutive units. However, once dissolved in ILs, the CH, being a charged polymer and having an ionic character, interacts with IL ions both electrostatically and through hydrogen bond formation. Such interactions can be envisaged from comparing the FTIR spectra of composites (Fig. 2, left) and composite ionogels (AG–CH; 50:50) (Fig. S2†) wherein a shifting of major peaks of the biopolymer is observed upon ionogel formation. The strength of composite ionogels (AG–CH; 50:50) was found to be even higher than the pure CH ionogel indicating the comparatively higher involvement of AG in the gelation process. The IL is immobilized in the gel matrix through strong hydrogen bonding between protons of imidazolium and oxygens of AG/CH as well as due to the hydrogen bonding between –OH/–NH$_2$ protons of AG/CH and Cl ions.

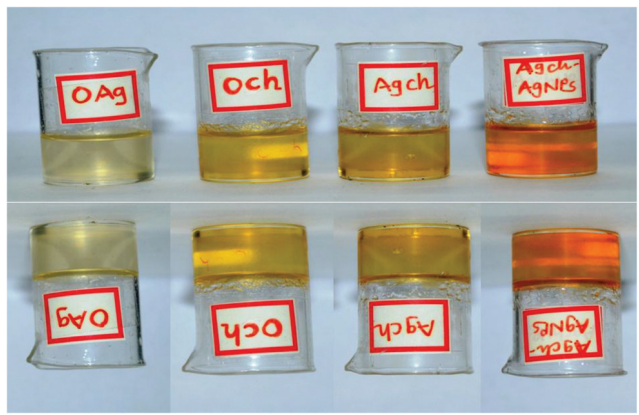

Fig. 5 Photographs of IL–biopolymer sols and ionogels (inverted); agarose (OAg), chitosan (OCh), agarose–chitosan composite (AgCh), and nanocomposite (AgCh–AgNPs).

Table 2 Gelling temperature ($T_{gel}$), melting temperature ($T_m$), gel strength and conductivity of ionogels ($\kappa_{ionogel}$)

| Composition | $T_{gel}$ (°C) | $T_m$ (°C) | Gel strength (g cm$^{-2}$) at 30 °C | $\kappa_{ionogel}$ (mS cm$^{-1}$) at 25 °C |
|---|---|---|---|---|
| AG | 40 | 53 | 140 | 0.065 |
| CH | 67 | 96 | 600 | 0.099 |
| AG–CH (50:50) | 56 | 84 | 670 | 0.075 |
| Ag NPs–AG–CH | 38 | 80 | 590 | 0.675 |

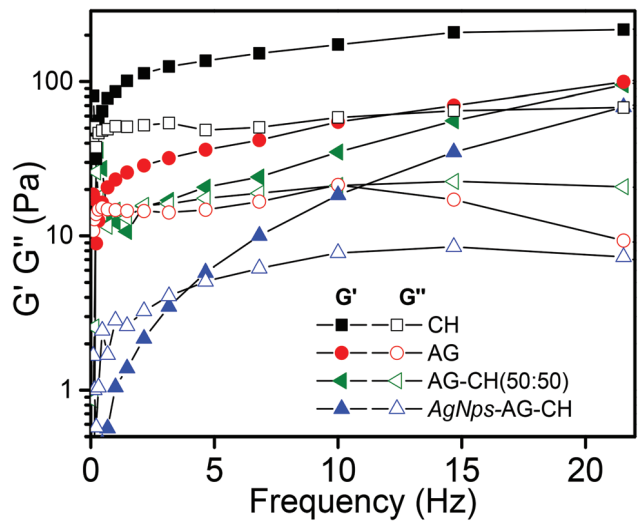

Fig. 6 Frequency dependence of dynamic storage (solid symbols) and loss moduli (open symbols) ($G'$ and $G''$) of different ionogels at a strain amplitude of $\gamma$ = 1% and 25 °C temperature.

Dynamic shear measurements were performed on various ionogels at 25 °C. The results of frequency dependences of dynamic storage ($G'$) and loss moduli ($G''$) are demonstrated in Fig. 6. In all the cases $G'$ is larger than $G''$, and is nearly frequency independent showing a solid-like behavior of the ionogels. Thermo-responsive sol–gel transition was observed from the temperature dependences of $G'$ and $G''$ measured at a frequency of 1 s$^{-1}$ (Fig. 7). A transition for both $G'$ and $G''$ as a function of temperature is indicative of melting temperature ($T_m$). Before the crossover of $G'$ and $G''$ IL–biopolymer solutions are optically transparent ionogels. $T_m$ obtained from temperature dependent rheology measurements are close to that observed by visual inspection reported in Table 2. We also conducted the dynamic strain sweep experiments on various ionogels over a wide range of strains (0.1 to 10%) at a frequency of 100 rad s$^{-1}$. The observed $G'$ and $G''$ trends of various ionogels under large strains are shown in Fig. 8. For most of the ionogels a linear viscoelastic regime is maintained indicating no deformation of the gel microstructure even at 10% strain at a frequency of 100 rad s$^{-1}$. In the case AG ionogels the $G'$ and $G''$ gap slightly decreases with the increase in strain showing a decrease in solid-like behaviour with a

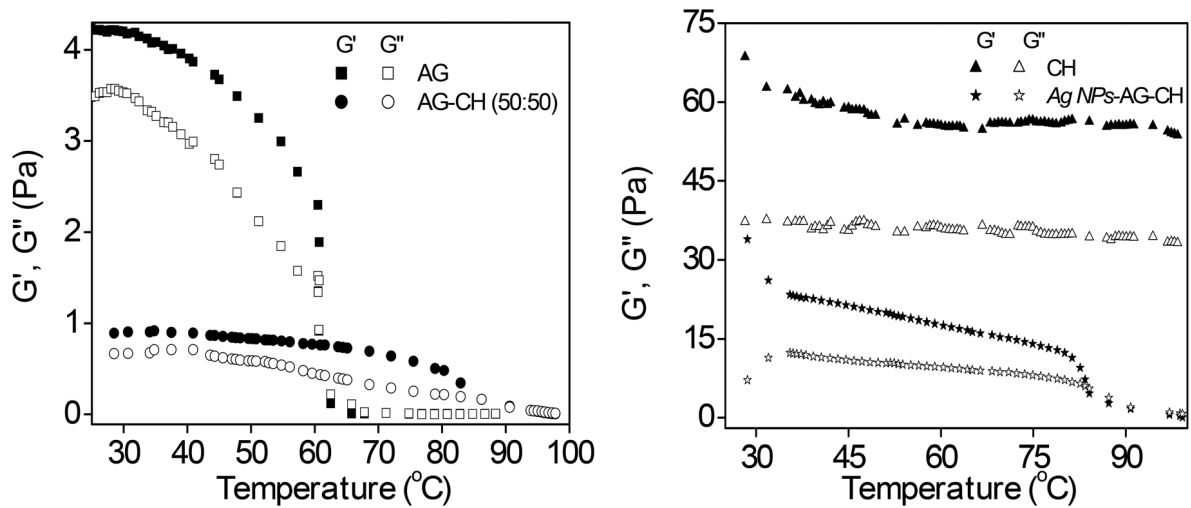

Fig. 7 Temperature dependence variation of dynamic storage (solid symbol) and loss moduli (open symbol) ($G'$ and $G''$) of ionogels; (left) AG and AG–CH (50 : 50) and (right) CH and Ag NPs–AG–CH at frequency $f = 1$ s$^{-1}$ and a strain amplitude of $\gamma = 1\%$.

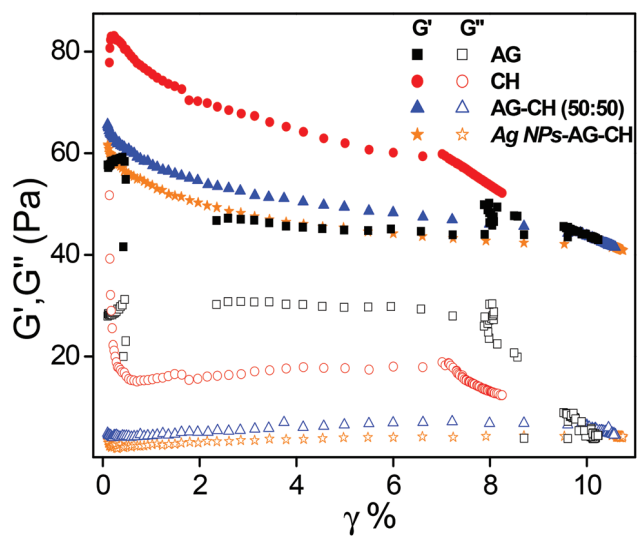

Fig. 8 Strain dependence of dynamic storage (solid symbol) and loss moduli (open symbol) ($G'$ and $G''$) of ionogels at selected frequency $\omega = 100$ rad s$^{-1}$.

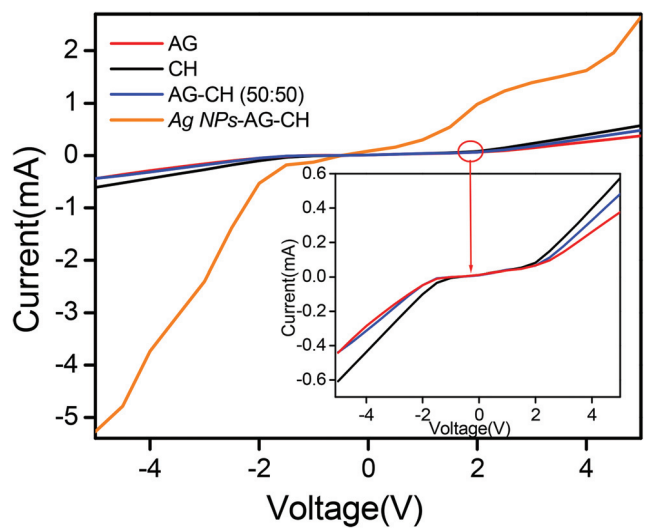

Fig. 9 (a) Current–voltage curves of AG, CH, AG–CH (50 : 50) and Ag NPs–AG–CH (50 : 50) ionogels.

continuous change in gel microstructure wherein the molecular connections in the gel network are disrupted under the large strains, while in CH and AG–CH composite ionogels, $G'$ remains quite higher than $G''$ within the region of linear viscoelasticity indicating the persistence of solid-like behavior. Other rheological properties of ionogels (tan δ and $\eta^*$ vs. temperature or frequency) are provided in ESI (Fig. S3†).

Since a low matrix of biopolymers (AG or CH) is enough to prepare ionogels of sufficiently good gel strength, such ionogels will have advantages in terms of retaining the conducting properties of native IL. Therefore, the conducting nature of ionogels was tested by current–voltage measurements. The steady state current–voltage profiles (Fig. 9) indicated that ionogels retained sufficiently good conducting properties,

CH-ionogels (0.099 mS cm$^{-1}$) being slightly more conducting than AG-ionogels (0.0653 mS cm$^{-1}$), whereas the conductivity of the AG–CH (50 : 50) composite ionogel (0.0750 mS cm$^{-1}$) lies between pure AG- and CH-ionogels showing homogeneous behavior of the composite material.

#### Ag NPs–AG–CH nanocomposite

Silver oxide nanoparticles (Ag NPs) were synthesized by heating 100 mM AgNO$_3$ in an AG–CH (50 : 50)–[C$_4$mim][Cl] solution at 100 °C under continuous stirring for 6 h. The solution colour change from yellow to light brownish confirmed the formation of silver nanoparticles *in situ*. Here, the biopolymers AG and CH played a dual role of both a stabilizer and a reducing agent. The characteristic surface plasmon resonance at 418 nm in UV-Vis spectra (Fig. 10) also confirmed the synthesis of Ag NPs. The addition of methanol in Ag NPs–AG–CH–[$C_4$mim][Cl] solution produced nanocomposites. While regenerating, the composite Ag NPs were well coated with AG–CH by interactions with $NH_2$ or OH groups of CH and AG similar to that observed for CH–alginate or CH–cellulose composites.[56,68] SEM images (Fig. 11a,b) of the nanocomposite show that Ag NPs are nearly spherical and uniformly decorated on the AG–CH surface. A TEM image of the nanocomposite (Fig. 11c) showed 10–20 nm size Ag NPs; the lattice fringes in high resolution and electron diffraction pattern (SAED) indicated their crystalline and polycrystalline nature. The crystalline nature of the nanocomposite is also confirmed by the comparison of XRD patterns of regenerated AG–CH and Ag NPs–AG–CH composites (Fig. 11d). AG–CH showed only a single broad band showing the amorphous nature of the composite, whereas the Ag NPs–AG–CH composite showed two intense peaks at 27.94 and 32.27, which correspond to (110) and (111) of $Ag_2O$. Apart from this, diffraction peaks at 46.34, 54.92 and 67.48 can be indexed to the (211), (220), (222) and (311) planes of face-centered cubic silver, respectively.

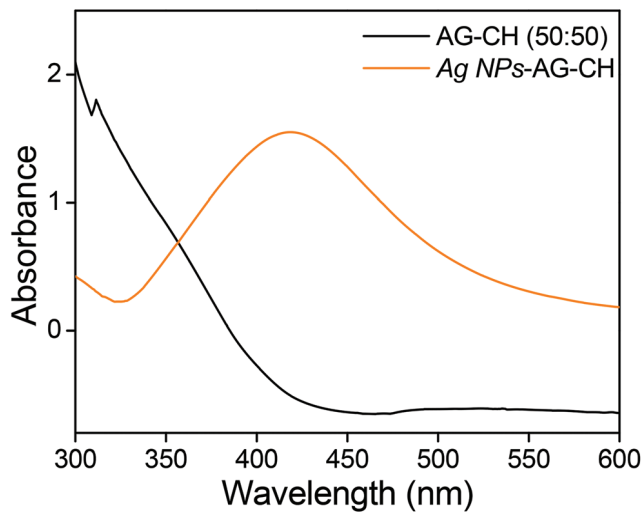

Fig. 10 UV-Vis spectra of AG–CH (50:50) IL solution and Ag NPs in AG–CH (50:50) IL solution.

**Ag NPs–AG–CH nanocomposite ionogels**

Hot solutions of 100 mM $AgNO_3$ in AG–CH (50:50)–[$C_4$mim][Cl] (at 100 °C, under continuous stirring for 6 h) upon cooling to room temperature resulted in the formation of homogeneous nanocomposite ionogels with well dispersed Ag NPs (the brownish colour of ionogels in Fig. 5 indicates the presence of Ag NPs). The inclusion of Ag NPs in ionogels reduced the gelling temperature and increased the melting temperature (Table 2) thus increasing the thermal hysteresis. Despite a longer period of heating in [$C_4$mim][Cl] (6 h), the nanocomposite ionogel showed a reasonably good gel strength (Table 2). The high gel strength of nanocomposite ionogels is possibly due to the bridging role of Ag NPs between polymer chains and forming a network structure along with the usual electrostatic and hydrogen bond interactions between polymer groups and IL ions similar to that observed by Shen et al.[69] for

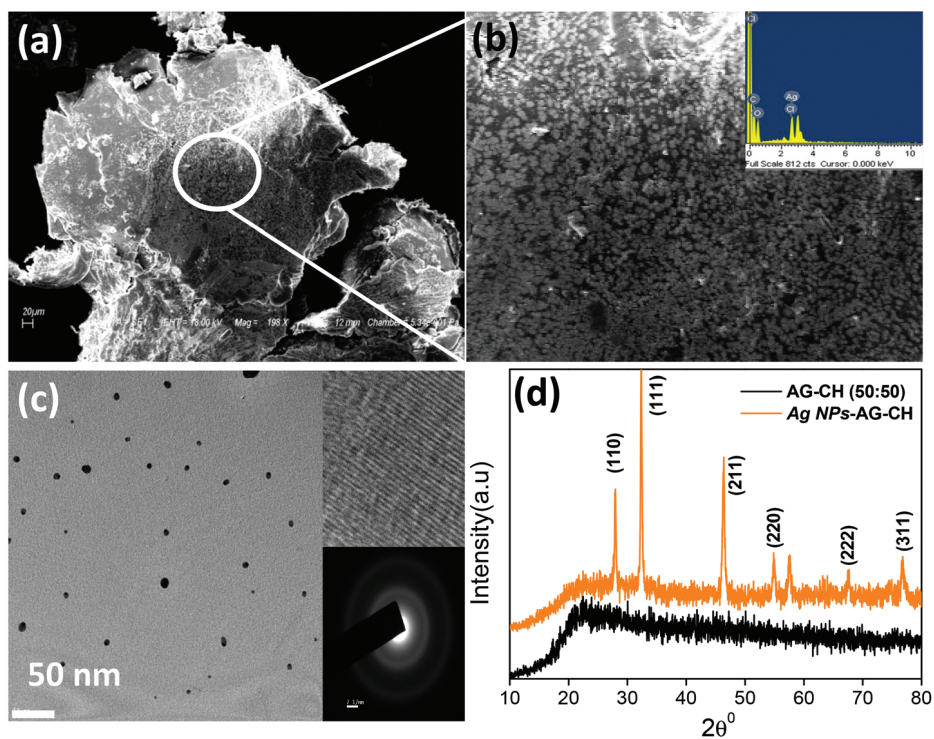

Fig. 11 (a, b) SEM and (c) TEM images of Ag NPs–AG–CH composite, and (d) comparison of XRD patterns of AG–CH and Ag NPs–AG–CH composite.

sodium carboxymethylcellulose (NaCMC) hydrogels where Ag(I) served as a bridging agent. The FTIR studies also indicated the interaction between composite and silver oxide nanoparticles in the ionogels (Fig. S2†). The FTIR spectrum of AG–CH ionogels showed characteristic IR bands of AG at 758, 852, and 949 cm$^{-1}$ for 3,6-anhydro-L-galactose skeletal bending, 1087 cm$^{-1}$ for –C–O–C and 1163 cm$^{-1}$ for glycosidic linkage and absorption bands of CH at 1642, 1564 and 1337 cm$^{-1}$ characteristic of amide I, amide II and amide III which are slightly shifted upon composite formation.[30,56] These FTIR peaks of AG–CH ionogels are further red shifted in the case of Ag NPs–AG–CH ionogels (Fig. S2†) indicating interaction of Ag NPs with the biopolymers.

Dynamic shear measurements of nanocomposite ionogels showed a different behavior than that observed for simple ionogels or composite ionogels (Fig. 6). Unlike other ionogels, in the case of nanocomposite ionogels, $G'$ increased continuously and $G''$ increased slightly initially and then became constant with the increase in frequency. At low frequency, $G' > G''$, demonstrating a viscous behavior, whereas $G''$ crosses $G'$ at ∼5 Hz and becomes much higher than $G'$ showing characteristics of cross linked gels. Temperature dependences of $G'$ and $G''$ measured at a frequency of 1 s$^{-1}$ (Fig. 7) showed a gel melting temperature ($T_m$) of ∼83 °C which is very close to that observed by visual inspection (Table 2). $G'$ and $G''$ tend to maintain a linear viscoelastic regime under the measured strains of 10% at a frequency of 100 rad s$^{-1}$ (Fig. 8) indicating no deformation of the gel microstructure. To test the suitability of the nanocomposite as a gel electrolyte, we determined the ionic conductivity from current–voltage measurements. The steady state current–voltage profile of Ag NPs in composite ionogels along with other ionogels is shown in Fig. 9. The conductivity of Ag NPs–AG–CH composite ionogels (0.675 mS cm$^{-1}$) was found to be an order of magnitude higher than that observed for normal and composite ionogels thus making them an attractive material for electronic devices.

## Conclusion

A viable one pot methodology for the preparation of composite/nanocomposite and nanocomposite ionogels of agarose and chitosan *via* simultaneous dissolution of biopolymers and *in situ* generation of silver oxide nanoparticles in the IL, [C$_4$mim][Cl], is reported. Homogeneous dissolution of biopolymers in the IL produced blends having a great compatibility of agarose and chitosan with strong hydrogen bonding between –OH/–NH$_2$ groups as was confirmed by FTIR and CD spectroscopy. Physiochemical characterization confirmed that the prepared composite materials have good stability and enhanced material properties compared with individual biopolymers. SEM, TEM, and XRD analyses showed decoration of the composites with 10–20 nm size highly crystalline silver oxide nanoparticles. Silver oxide nanoparticles and chitosan display antimicrobial activity; therefore, the nanocomposite materials produced would also be suitable for applications such as food packaging, wound dressing, drug delivery, *etc.* On the other hand, uniformly dispersed silver oxide nanoparticles–agarose–chitosan–ionic liquid gels of high mechanical strength and conducting properties were prepared. Such nanocomposite ionogels will be promising materials for quasi-solid dye sensitized solar cells, actuators, sensor based materials or electronic devices.

## Acknowledgements

The financial support provided by the Department of Science and Technology (DST), Government of India (project no. SR/S/PC-04/2010) is highly acknowledged. We thank the Centralized Instrumental Facility of our Institute for assistance.

## References

1 O. J. Cayre, S. T. Changand and O. D. Velev, *J. Am. Chem. Soc.*, 2007, **129**, 10801–10806.
2 D. B. Saris, N. Mukherjee, L. J. Berglund, F. M. Schulz and S. W. O'Driscoll, *Tissue Eng.*, 2000, **6**, 531–537.
3 R. Muzzarelli, G. Biagini, A. Pugnaloni, O. Fillippini, V. Baldassarre, C. Castaldini and C. Rizzoli, *Biomaterials*, 1989, **10**, 598–603.
4 R. Muzzarelli, V. Baldassarre, F. Conti, P. Ferrara, G. Biagini, G. Gazzanelli and C. Rizzoli, *Biomaterials*, 1988, **9**, 247–252.
5 J. A. Hubbell, *Biotechnol.*, 1998, **3**, 565–576.
6 M. F. Nazarudin, A. A. Shamsuri and M. N. Shamsudin, *Int. J. Pure Appl. Sci. Technol.*, 2011, **3**, 35–43.
7 S. H. Teng, P. wang and H. Kim, *Mater. Lett.*, 2009, **63**, 2510–2512.
8 E. A. El-hafian, M. M. A. Nasef, A. H. Yahayam and R. A. Khan, *J. Chil. Chem. Soc.*, 2010, **55**, 130–136.
9 *Green Industrial Applications of Ionic Liquids*, ed. R. D. Rogers, K. R. Seddon and S. Volkov, NATO Science Series, Kluwer, DordreCH, 2002.
10 R. D. Rogers and K. R. Seddon, *Ionic Liquids as Green Solvents; ACS Symposium Series 856*, American Chemical Society, Washington, DC, 2003.
11 P. Wasserscheid and T. Welton, *Ionic Liquids in Synthesis*, Wiley-VCH; Verlag GmbH & Co. KGaA, Germany, 2nd edn, 2008.
12 M. A. P. Martins, C. P. Frizzo, D. N. Moreira, N. Zanatta and H. G. Bonacorso, *Chem. Rev.*, 2008, **108**, 2015–2050.
13 R. Bogel-Łukasik, N. M. T. Lourenço, P. Vidinha, M. D. R. Gomes da Silva, C. A. M. Afonso, M. Nunes da Ponte and S. Barreiros, *Green Chem.*, 2008, **10**, 243–248.
14 U. Domańska and R. Bogel-Łukasik, *J. Phys. Chem. B*, 2005, **109**, 12124–12132.
15 R. P. Swatloski, S. K. Spear, D. John, J. D. Holbrey and R. D. Rogers, *J. Am. Chem. Soc.*, 2002, **124**, 4974–4975.

16 I. Kilpeläinen, H. Xie, A. King, M. Granstrom, S. Heikkinen and D. S. Argyropoulos, *J. Agric. Food Chem.*, 2007, **55**, 9142–9148.
17 M. Zavrel, D. Bross, M. Funke, J. Buchs and A. C. Spiess, *Bioresour. Technol.*, 2009, **100**, 2580–2587.
18 A. Brandt, J. P. Hallett, D. J. Leak, R. J. Murphy and T. Welton, *Green Chem.*, 2010, **12**, 672–679.
19 N. Sun, H. Rodríguez, M. Rahman and R. D. Rogers, *Chem. Commun.*, 2011, **47**, 1405–1421.
20 S. S. Y. Tan and D. R. MacFarlane, *Top. Curr. Chem.*, 2010, **290**, 311–339.
21 M. E. Zakrzewska, E. Bogel-Łukasik and R. Bogel-Łukasik, *Energy Fuels*, 2010, **24**, 737–745.
22 A. Brandt, J. Gräsvik, J. P. Hallett and T. Welton, *Green Chem.*, 2013, **15**, 550–583.
23 H. Xie, S. Zhang and S. Li, *Green Chem.*, 2006, **8**, 630–633.
24 Y. S. Wu, T. Sasaki, S. Irie and K. Sakurai, *Polymer*, 2008, **49**, 2321–2327.
25 W.-T. Wang, J. Zhu, X.-Li Wang, Y. Huang and Y.-Z. Wang, *J. Macromol. Sci., Part B: Phys.*, 2010, **49**, 528–541.
26 Y. Qin, X. Lu, N. Sun and Robin D. Rogers, *Green Chem.*, 2010, **12**, 968–971.
27 Q. Chen, A. Xu, Z. Li, J. Wang and S. Zhangb, *Green Chem.*, 2011, **13**, 3446–3452.
28 P. S. Barber, C. S. Griggs, J. R. Bonnerc and R. D. Rogers, *Green Chem.*, 2013, **15**, 601–607.
29 L. Liu, S. Zhou, B. Wang, F. Xu and R. Sun, *J. Appl. Polym. Sci.*, 2013, **129**, 28–35.
30 T. Singh, T. J. Trivedi and A. Kumar, *Green Chem.*, 2010, **12**, 1029–1035.
31 T. J. Trivedi, D. N. Srivastava, R. D. Rogers and A. Kumar, *Green Chem.*, 2012, **14**, 2831–2839.
32 X. Sun, B. Peng, Y. Ji, J. Chen and D. Li, *AIChE*, 2009, **55**, 2060–2069.
33 W. Xiao, Q. Chena, Y. Wua, T. Wua and L. Daib, *Carbohydr. Polym.*, 2011, **83**, 233–238.
34 H. Ma, B. S. Hsiao and B. Chu, *Polymer*, 2011, **52**, 2594–2599.
35 C. Stefanescua, W. H. Dalya and I. I. Negulescu, *Carbohydr. Polym.*, 2012, **87**, 435–443.
36 S. S. Silva, T. C. Santos, M. T. Cerqueira, A. P. Marques, L. L. Reys, T. H. Silva, S. G. Caridade, J. F. Mano and R. L. Reis, *Green Chem.*, 2012, **14**, 1463–1470.
37 A. A. Shamsuri and R. Daik, *Materials*, 2013, **6**, 682–698.
38 J. L. Bideau, L. Viaub and A. Vioux, *Chem. Soc. Rev.*, 2011, **40**, 907–925 and references therein.
39 K. Suzuki, M. Yamaguchi, M. Kumagai, N. Tanabe and S. Yanagida, *C. R. Chim.*, 2006, **9**, 611–616.
40 J.-i. Kadokawa, M.-a. Murakami and Y. Kaneko, *Carbohydr. Res.*, 2008, **343**, 769–772.
41 P. K. Singh, B. Bhattacharya, R. K. Nagarale, K.-W. Kim and H.-W. Rhee, *Synth. Met.*, 2010, **160**, 139–142.
42 N. Wang, X. Zhang, H. Liu and B. He, *Carbohydr. Polym.*, 2009, **76**, 482–484.
43 A. A. Shamsuri, D. K. Abdullah and R. Daik, *Cell. Chem. Technol.*, 2012, **46**, 45–52.
44 J.-i. Kadokawa, M.-a. Murakami, A. Takegawa and Y. Kaneko, *Carbohydr. Polym.*, 2009, **75**, 180–183.
45 K. Prasad, Y. Kaneko and J.-i. Kadokawa, *Macromol. Biosci.*, 2009, **9**, 376–382.
46 P. Vidinha, N. M. T. Lourenço, C. Pinheiro, A. R. Brás, T. Carvalho, T. Santos-Silva, A. Mukhopadhyay, M. J. Romão, J. Parola, M. Dionisio, J. M. S. Cabral, C. A. M. Afonsoc and S. Barreiros, *Chem. Commun.*, 2008, 5842–5844.
47 K. Prasad, H. Izawa, Y. Kaneko and J.-i. Kadokawa, *J. Mater. Chem.*, 2009, **19**, 4088–4090.
48 N. M. Sangeetha, S. Bhat, G. Raffy, C. Belin, A. Loppinet-Serani, C. Aymonier, P. Terech, U. Maitra, J. P. Desvergne and A. D. Guerzo, *Chem. Mater.*, 2009, **21**, 3424–3432.
49 J. Zhang, B. Zhao, L. Meng, H. Wu, X. Wang and C. Li, *J. Nanopart. Res.*, 2007, **9**, 1167–1171.
50 C. S. Love, V. Chechik, D. K. Smith, K. Wilson, I. Ashworth and C. Brennan, *Chem. Commun.*, 2005, 1971; H. Basit, A. Pal, S. Sen and S. Bhattacharya, *Chem.–Eur. J.*, 2008, **14**, 6534–6545.
51 (*a*) W.-X. Li, C. Stampfl and M. Scheffler, *Phys. Rev. B: Condens. Matter*, 2003, **68**, 165412–165427; (*b*) B. E. Breyfogle, C. Hung, M. G. Shumsky and J. A. Switzer, *J. Electrochem. Soc.*, 1996, **143**, 2741–2746; (*c*) Y. Ida, S. Watase, T. Shinagawa, M. Watanabe, M. Chigane, M. Inaba, A. Tasaka and M. Izaki, *Chem. Mater.*, 2008, **20**, 1254–1256.
52 (*a*) F. Derikvand, F. Bigi, R. Maggi, C. G. Piscopo and G. J. Sartori, *J. Catal.*, 2010, **271**, 99–103; (*b*) W. Wang, Q. Zhao, J. Dong and J. Li, *Int. J. Hydrogen Energy*, 2011, **36**, 7374–7380.
53 V. V. Petrov, T. N. Nazarova, A. N. Korolev and N. F. Kopilova, *Sens. Actuators, B*, 2008, **133**, 291–295.
54 Z. Hu, W. L. Chan and Y. S. Szeto, *J. Appl. Polym. Sci.*, 2008, **108**, 52–56.
55 S. Tripathi, G. K. Mehrotra and P. K. Dutta, *Bull. Mater. Sci.*, 2011, **34**, 29–35.
56 S. Sharma, P. Sanpui, A. Chattopadhyay and S. S. Ghosh, *RSC Adv.*, 2012, **2**, 5837–5843.
57 K. Chou and Y. Lai, *Mater. Chem. Phys.*, 2004, **83**, 82–88.
58 H. Huang, Q. Yuan and X. Yang, *Colloids Surf., B*, 2004, **39**, 31–37.
59 P. Sanpui, A. Murugadoss, P. V. D. Prasad, S. S. Ghosh and A. Chattopadhyay, *Int. J. Food Microbiol.*, 2008, **124**, 142–146.
60 M. K. Shukla, R. P. Singh, C. R. K. Reddy and B. Jha, *Bioresour. Technol.*, 2012, **107**, 295–300.
61 J. S. Craigie and C. Leigh, Carrageenans and agars, in *Handbook of phycological methods*, ed. J. A. Hellebust and J. S. Craigie, Cambridge University Press, Cambridge, 1978, pp. 109–131.
62 J. Brugnerottoa, J. Lizardib, F. M. Goycooleab, W. ArguÈelles-Monalc, J. DesbrieÁresa and M. Rinaudoa, *Polymer*, 2001, **42**, 3569–3580.
63 D. Christiaen and M. Bodard, *Bot. Mar.*, 1983, **26**, 425–427.

64 K. Prasad, G. Mehta, R. Meena and A. K. Siddhanta, *J. Appl. Polym. Sci.*, 2006, **102**, 3654–3663.
65 S.-H. Teng, P. Wang and H.-E. Kim, *Mater. Lett.*, 2009, **63**, 2510–2512.
66 W. Feng, Y. Li and P. Ji, *AIChE*, 2012, **58**, 285–291.
67 M. K. Cheung, K. P. Y. Wan and P. H. Yu, *J. Appl. Polym. Sci.*, 2002, **86**, 1253–1258.
68 Z. Liu, H. Wang, B. Li, C. Liu, Y. Jiang, G. Yu and X. Mu, *J. Mater. Chem.*, 2012, **22**, 15085–15091.
69 J.-S. Shen and B. Xu, *Chem. Commun.*, 2011, **47**, 2577–2579.